\documentclass[12pt]{article}
\usepackage{amsmath}
\usepackage{amssymb}
\usepackage{epsfig,hyperref}
\usepackage{color,colordvi}
\def\colour4colour#1{\Blue{#1}}

\newcommand{\eq}{\begin{equation}}
\newcommand{\eqx}{\end{equation}}
\newcommand{\eqn}{\begin{eqnarray}}
\newcommand{\bi}{\begin{itemize}}
\newcommand{\eqnx}{\end{eqnarray}}
\newcommand{\ei}{\end{itemize}}

\begin{document}

\begin{titlepage}


\vspace{0.5cm}

\begin{center}

\huge
{A bound on the effective gravitational coupling from semiclassical black holes }

\vspace{0.8cm}

\large{R. Brustein$^{1}$, G. Dvali$^{2,3,4}$, G. Veneziano$^{2,5,6}$}

\normalsize

\vspace{0.5cm}

{\sl $^1$Department of Physics, Ben-Gurion
University,\\
    Beer-Sheva, 84105 Israel}

\vspace{.1in}

{\sl $^2$ CERN, Theory Unit, Physics Department, \\ CH-1211 Geneva 23, Switzerland}

\vspace{.1in}

{\sl $^3$  CCPP, Department of Physics, New York University\\
4 Washington Place, New York, NY 10003 }
\vspace{.1in}

{\sl $^4$  Max-Planck-Institute for Physics,\\
F\"ohringer  Ring 6, D-80805, M\"unchen,  Germany}

\vspace{.1in}

{\sl $^5$Coll\`ege de France, 11 Place M. Berthelot, 75005 Paris, France}

\vspace{.1in}

{\sl $^6$Galileo Galilei Institute and Istituto Nazionale di Fisica Nucleare, Florence, Italy}

\begin{abstract}
\noindent
 We show that the existence of semiclassical black holes of size as small as a minimal length scale $l_{UV}$  implies a bound  on a gravitational analogue of  't-Hooft's coupling  $\lambda_G(l)\equiv N(l) G_N/l^2$ at all scales $l \ge l_{UV}$.   The proof is valid for any metric theory of gravity that consistently extends Einstein's gravity and is based on two assumptions about semiclassical black holes: i) that they emit as black bodies, and ii) that they are perfect quantum emitters. The examples of higher dimensional gravity and of weakly coupled string theory are used to explicitly check our assumptions and to verify that the proposed bound holds. Finally, we discuss  some consequences of the bound for  theories of quantum gravity in general and for string theory in particular.

\end{abstract}
\end{center}

\end{titlepage}

\newpage

\section{Introduction}

The prevailing common wisdom is that Einstein's gravity --together with a quantum-field-theory of matter-- is only an effective, large-distance description of physics that requires a consistent ultraviolet completion. Any such completion (as provided, for example, by  string theory) will modify physics below a length scale $l_{UV}$, the ultraviolet (UV) cutoff.    The goal of this paper is to prove a universal upper bound on the dimensionless effective gravitational coupling for all scales larger than the cutoff scale in all such theories of quantum gravity.

Let us define the gravitational analogue of  't-Hooft's coupling\footnote{For the sake of clarity our discussion refers to four spacetime dimensions, but our results can be straightforwardly extended  to a general number $D$ of spacetime dimensions.}
\begin{equation}
\label{deflam}
\lambda_G(l) = N(l) \frac{l_P^2}{l^2},
\end{equation}
where $l_P$ is the Planck length\footnote{We use units in which $c,\hbar, k_{B}=1$ and neglect purely numerical factors throughout the paper.}  $l_P = G_N^{1/2}$ and $N(l)$ is the number of light species at the scale $l$ (to be defined precisely later).
We will show, under some further mild assumptions to be specified below, that at the cutoff scale $\lambda_G(l_{UV}) < 1$ and hence that
\begin{equation}
\label{bound}
\lambda_G(l) < 1\ \text{for}\ l \ge l_{UV}.
\end{equation}
This is the main result of our paper.

The bound  (\ref{bound}) has appeared previously in several contexts \cite{Gab}, \cite{GV01}, \cite{Dvali, cesar}. In \cite{Gab} and \cite{GV01} the relation (\ref{bound}) was introduced  in the context of perturbative renormalization of the graviton kinetic term  by species loops.   Barring possible cancelations,  this contribution is  proportional to the number of species, suggesting that in a theory with many species there is a natural hierarchy between the Planck mass and the cutoff.

The  derivation of \cite{Dvali,cesar} was based on using non-perturbative BH arguments, showing that BHs obeying certain well-defined conditions of semiclassicality (to be elaborated  below)  cannot exist beyond the scale
\begin{equation}
\label{defl*}
l_{SCBH} \equiv l_P \sqrt{N(l_{SCBH})} \, ,
\end{equation}
and then asserting that $l_{SCBH}$ bounds $l_{UV}$, thus implying the bound (\ref{bound}).
The key point of \cite{Dvali,cesar}  is this connection between  $l_{SCBH}$  and   $l_{UV}$.
However, the analysis was limited by the class of theories in which classical gravity at short-distances never becomes weaker than Einsteinian gravity.  One of the novelties in our case is, that we show that any other situation
is inconsistent (i.e., short-distance gravity can never be weaker than Einsteinian gravity) and thus  the bound in absolute. The scale $l_{SCBH}$ was first discussed in \cite{Brustein:2000hh} in the context of a different bound on the number of species from vacuum stability. There the scale $l_{SCBH}$ was used to find the range of validity of the semiclassical arguments.

To the best of our knowledge bounds of the form (\ref{bound}) were first discussed in connection with cosmological entropy bounds \cite{Bekenstein:1989wf} where it was suggested that the temperature $T$ of a radiation dominated universe is universally bounded $T^2<M_P^2/N$, where $M_P=1/l_P$ is the Planck mass. The proposal was then extended to a fixed region at temperature $T$ in \cite{Brustein:2004wb}. Previously, using  the Generalized second law it was shown that the scalar curvature $R$ satisfies a similar bound  $R < M_P^2/N$ in Einstein gravity \cite{Brustein:1999ua} and in string theory \cite{Brustein:1999ay}.

The central question that we investigate and answer in the current work  is  whether a sensible consistent {\it classical} modification of Einstein gravity could exist that would allow semiclassical BHs whose size is smaller than  $l_{SCBH}$.   If any such modification would exist, it would  imply the existence of a  new semiclassical  gravitational regime beyond the scale $l_{SCBH}$.  If this were so
(\ref{bound}) would not be universal, rather it would bound the scale of new gravitational physics.
In the present paper we show that (\ref{bound}) is universal.  We first show, that  $l_{SCBH}$ is an absolute lower bound on the size of semiclassical BHs in {\it any} consistent theory of gravity. We then show that  in any consistent theory of gravity $l_{SCBH}< l_{UV}$.  Adding some reasonable assumptions about the dependence on $l$ of $N(l)$ we prove the bound (\ref{bound}) in its  full generality.

\section{Assumptions on semiclassical black holes}

Let us consider neutral static and non-rotating BHs. They can be described in terms of three parameters: the mass $M$, the Schwarzschild radius $R_S$ and the inverse temperature $\beta=1/T$.  In Einstein gravity these three parameters are related in a simple way. While the existence of such relations is guaranteed by the no-hair theorem \cite{nohair}, we will leave their exact form unspecified in order to allow for possible modifications of Einstein's gravity above a certain energy scale.

Following \cite{Dvali} let us define semiclassical BHs as those satisfying the following intuitive physical conditions. That the BH size and inverse temperature decrease at a speed slower than the speed of light $(a)\ - \frac{dR_{S}}{dt} < 1,\
(b)\ - \frac{d\beta}{dt} < 1$; That the fractional change of the mass of the BH be small during both the thermal and the light crossing time scales,
$(c)\
-\frac{R_S}{M} \frac{dM}{dt}< 1,
\ (d)\
-\frac{\beta}{M} \frac{dM}{dt}< 1$ and  that the BH be metastable
$(e)\
\frac{\Gamma}{M} < 1.
$
Here the definition of $\Gamma$ is the same as for the elementary species: the inverse of the time it takes for the first transition to a lower energy state via the emission of a light quantum.

We make two basic assumptions about the nature of semiclassical BHs:

i) that they emit as black bodies, so that:
\begin{equation}
\label{BB}
-\frac{dM}{dt}= N(\beta) \beta^{-4} R_S^2 \, .
\end{equation}
Here $N(\beta)$ is the number of light species into which the BH can decay. We have ignored numerical grey factors and additional numerical factors related to the statistics of the species. None of the decay channels of the BH are expected to be parametrically suppressed at energies $\sim 1/\beta$ which is the main energy range of the BH emission.   Since we assume that the BHs are black bodies it follows that  $N(\beta)$ is equal to the number of species that can be in thermal equilibrium at (inverse) temperature $\beta$.

ii)  that they are perfect quantum emitters i.e. that they cannot either emit any particles classically nor can their emission be controlled classically. This implies that the (quantum) wavelength of the particles they emit is not smaller than  $R_S$. Since, according to our first assumption, the semiclassical BHs radiate like  black bodies, this implies that $R_S^{-1}$, being the energy of the emitted quanta, also  bounds the BH temperature i.e.
\begin{equation}
\label{trs}
 R_S/\beta \le1.
\end{equation}
We will show that for a neutral {\it classically-static} non-rotating BHs the above inequality is saturated.
\footnote{ The assumption of classical time independence is important
for our analysis. Otherwise, for microscopic semiclassical BHs that are localized in compact extra dimensions the condition (\ref{BB}) can be easily violated \cite{Dvali}. See Sect~(\ref{extraD}) for further discussion of this point.}

We can now substitute Eqs.~(\ref{BB}), (\ref{trs}) into inequalities~$(a)-(d)$ and, after some simple algebra,  obtain the two following inequalities
\begin{eqnarray}
\label{bounds1a}
R_S M &>& N(\beta)\ \frac{d\ln R_S}{d \ln M} \\
\label{bounds1b}
R_S M &>& N(\beta).
\end{eqnarray}
In the following section we show that they reduce to the single inequality~(\ref{bounds1b}).
Finally, the inequality~$(e)$ is also implied by the above one. Indeed, for a black body, $-dM/dt= \Gamma T$ so $\Gamma/M<1$ is equivalent to inequality~$(d)$.

\section{A bound on the effective gravitational coupling}

\subsection{Einstein gravity}
\label{EGefflambda}

Let us consider for the moment the case of Einstein gravity with a constant and fixed number of light metastable  species $N$ (to be defined more precisely below). In this case  $ M=M_P^2 R_S $ and $R_S/\beta=1$.
Then inequalities~(\ref{bounds1a}) and~(\ref{bounds1b}) are equivalent and imply that
\begin{equation}
\label{Ebound}
R_S > l_P \sqrt{N} = l_{SCBH}
\end{equation}
and using the definition~(\ref{deflam}) of $\lambda_G$
\begin{equation}
\label{Lbound}
\lambda_G(l_{SCBH}) <1.
\end{equation}

Additionally, the bound~(\ref{Ebound}) implies that semiclassical BHs of size smaller than $l_{SCBH}$ cannot exist. We will show below that gravity can no longer be treated as weakly coupled below that length scale. Thus the effective description that we have used breaks down and the scale $l_{SCBH}$ should be considered as a lower bound on the actual short distance cutoff of the theory:
\begin{equation}
\label{UVbound}
l_{UV} \ge l_{SCBH}.
\end{equation}
Combining (\ref{UVbound}) with (\ref{Lbound}) and since $\lambda_G(l)/\lambda_G(l_{UV})=l_{UV}^2/l^2$  we obtain
\begin{equation}
\label{boundI}
\lambda_G(l_{UV}) <1.
\end{equation}
From this inequality it follows that
$\lambda_G(l) <1$ for $l\ge l_{UV}$, which is the result announced in Eq.~(\ref{bound}).

We will prove  bound~(\ref{UVbound}) by showing that the opposite assumption  $l_{UV} < l_{SCBH}$ leads to a contradiction. For this we generalize the two-observer thought experiment described in \cite{cesar} where
a collapsing distribution of matter, for example, dust, was considered.  Let the total mass of this distribution be $M$ and a corresponding (would-be) Schwarzschild radius be $R_S > l_{SCBH}$.
If $l_{UV} < \l_{SCBH}$  it is possible to prepare an initial distribution of matter which would cross into its own Schwarzschild radius $R_S$ while the curvature is smaller than $1/ l_{UV}^2$. Now consider two observers,  one (Alice) is observing the collapse from the far away while the other (Bob) is a freely-falling with the collapsing matter. The equivalence principle requires that Alice and Bob should agree on the fate of the matter distribution if they are in causal contact with each other. Bob can continuously monitor the matter density and the curvature, by measuring the tidal forces,  and finds that the curvature and the matter density are always small so quantum corrections to any classical process are small.
Alice, on the other hand, sees a violent decay of a quantum mechanical object over a time scale shorter than $R_S$. Since Alice and Bob remain in causal contact during the collapse because a BH horizon does not have time to form, they can compare their results and verify their disagreement on the fate of the collapsing matter.
The issue of whether or not during his uninterrupted classical journey Bob would eventually end up in a singularity is irrelevant to our argument since Bob and Alice have enough time to compare their observations before any high curvature region is formed.

Let us now allow $N$ to depend on $l$. To discuss this case we need to define $N$ more precisely. We wish to consider theories that at an energy scale $\Lambda=1/l$ have a finite number $N(l)$ of light species whose mass is smaller than $\Lambda$, $m<\Lambda$ and whose decay width is smaller than their mass $\Gamma<m$. Of particular interest is the case that the energy scale $\Lambda$ is the UV cutoff scale $\Lambda_{UV}=1/l_{UV}$. The number $N(l)$ includes the graviton and possibly other gravitational degrees of freedom and the decay width $\Gamma$ is defined as the inverse of the lifetime of the state which is the time that it takes for the first transition to a lower energy state via the emission of a light quantum. We shall assume that the coupling of all the light species $N(l)$ is such that they can be at thermal equilibrium at (inverse) temperature $\beta=l$.
We shall consider only metric theories of gravity
and define the scale $l_{UV}$ for such theories as the scale above which exchanges of metric perturbations in elementary particle processes become strong. Obviously, it follows that for curvatures less than $1/l_{UV}^2$ gravity is weak and semiclassical. It may well be that Einstein's gravity is modified for scales well above $l_{UV}$, for example,  if large extra dimensions of size $R > l_{UV}$ exist.

Allowing $N$ to depend on $l$ we have $\frac{\lambda_G(l)}{\lambda_G(l_{UV})}=\frac{N(l)}{N(l_{UV})} \frac{l_{UV}^2}{l^2}$. To prove the bound in this case we need to add the reasonable assumption  that the ratio $N(l)l_P^2/l^2$ is maximal at the highest scale $l=l_{UV}$. In other words, we assume that the unlikely possibility that  $N(l)$ grows faster than $(l/l_{UV})^2$ in the infrared is not realized. From inequality~(\ref{boundI}) and the assumption that  that $N(l_{UV})l_P^2/l_{UV}^2$ is maximal it follows that
$\lambda_G(l) <1$ for $l\ge l_{UV}$, as in the case for constant $N$.

\subsection{Extensions of Einstein gravity}

Let us consider a general theory of gravity which is not necessarily Einstein's. We wish to show that the bound~(\ref{Lbound}) holds for any such extension provided it is a consistent one.

We will do so by showing that, for a given Schwarzschild radius $R_S$, the BH mass $M$ and thus the product $M R_S$ is maximal in Einstein gravity.  This implies that  $l_{SCBH}$ is  the shortest scale for any consistent semiclassical description  of gravity.  Since, for a given $M$, the Schwarzschild radius $R_S$ is the largest in Einstein gravity,  the proof of bound~(\ref{UVbound}) by the two-observers thought experiment remains valid too. In  other words, we will show that an observer attempting to determine the BH mass by measuring a free-fall acceleration of a probe source can only detect a stronger acceleration than the one she would detect at the same distance if the theory were Einstein gravity.

In order to quantify  this argument, let us consider a  metric perturbation about  flat spacetime,
$
g_{\mu\nu} \, = \, \eta_{\mu\nu} \, + \, h_{\mu\nu}
$.
This metric perturbation is sourced by  the BH. The key point is that outside  the BH horizon gravity is  weakly coupled.  Thus, the leading  gravitational process contributing to the acceleration  of  the probe source in the  one-particle exchange amplitude.
In a generic weakly-coupled theory of gravity,   $h_{\mu\nu}$ can be decomposed into the spin-2 and spin-0 states. Other spins do not contribute at the linear level, due to the conservation of the source, and are therefore  irrelevant.
The one particle exchange amplitude among the BH and the probe,
$
G \, \equiv \,  t^{\mu\nu} \langle h_{\mu\nu} h_{\alpha\beta}\rangle T^{\alpha\beta}
$
can be decomposed into irreducible representations as follows,
\begin{equation}
\label{decomp}
 G \, = \, {1 \over M_P^2} {t_{\mu\nu}T^{\mu\nu} \, - {1 \over 2} t_{\mu}^{\mu} T_{\nu}^{\nu}
\over p^2}\,
+ \, \sum_i \,  {1 \over M_i^2} {t_{\mu\nu}T^{\mu\nu} \, - \, {1 \over 3} t_{\mu}^{\mu} T_{\nu}^{\nu}
\over p^2 \, - \, m_i^2}\, + \,  \sum_j {1 \over ({\overline{M}}_j)^2} {t_ {\mu}^{\mu} T_{\nu}^{\nu}
\over p^2 \, - \, ({\overline{m}}_j)^2}\, ,
\end{equation}
where we have explicitly separated the massless spin-2 (two physical polarizations) massive spin-2
(five physical polarizations) and spin-0 contributions respectively.   $M_i$, $(\overline{M}_j)$  are
the coupling strengths and $m_i$, $\overline{m}_j$ are the masses of spin-2 and spin-0 states respectively.

Equation~(\ref{decomp}) is the most general ghost-free structure for the exchange
between the conserved energy-momentum sources, which in addition requires that all the coefficients are positive \cite{ghosts}. Then all the interactions in Eq.~(\ref{decomp}) are attractive, so they  cannot induce mass screening that reduces the acceleration of the probe. It is also clear that a possible running of the coupling constants  $M_P(p^2)$, $M_i(p^2)$, $\overline{M}_j(p^2)$ or masses
$m_i^2(p^2)$, $\overline{m}_j^2(p^2)$
with the momentum (or distance) cannot change this conclusion,  since at any scale the decomposition~(\ref{decomp}) should be valid as a ghost-free spectral representation \cite{ghosts}. For example,  a mass-screening of a gravitating source would take place if the strength of the first term in~(\ref{decomp}) decreases in UV. This is impossible in a ghost free theory. If the spectral function $\rho(s)$ is positive-definite   the spectral representation  of the scalar  propagator
${1 \over M(p^2) p^2} \, = \, \int \, ds {\rho(s) \over p^2 - s}
$
mandates that $M^2(p)$ can only decrease in the UV.

We may now evaluate $h_{\mu\nu}(r)$, taking into account all the additional modes that can contribute to this exchange
$
h_{00}(r)=-\frac{M}{M_P^2}\frac{1}{r} \left(1 +\int\limits_0^\infty dm \rho(m) e^{-m r}\right)$.
As we have discussed, the spectral function $\rho(m)$ is positive definite for a ghost-free theory and by choosing the lower limit of the integral $I(r)=\int\limits_0^\infty dm \rho(m) e^{-m r}$ we have implemented the requirement that all masses are positive.

In this approximation the position of the horizon $R_S$ is approached for $h_{00}(R_S)=-1$,
\begin{equation}
\label{mrs}
M=M_P^2 \frac{R_S}{1+I(R_S)} \, .
\end{equation}
On the other hand, the temperature, in this approximation, is given by $T=d h_{00}/dr|_{r=R_S}$, so that, in agreement with Eq. (\ref{trs}),
\begin{equation}
\label{trsgen}
T R_S = 1+ \frac{\left|\frac{R_Sd I(R_S)}{dR_S}\right|}{1+I(R_S)} \sim 1\, ,
\end{equation}
where we have used that $I(R_S)$ is cutoff at $m = R_S^{-1}$.
In order to prove that the bound~(\ref{bounds1b}) is sufficient  we need to bound the factor $\frac{d\ln R_S}{d\ln_M}$ in Eq.~(\ref{bounds1a}).
From Eq.~(\ref{mrs}) we find
\begin{equation}
\frac{1}{M_P^2} \frac{dM}{dR_S}=\frac{1}{1+I(R_S)}\left( 1+\frac{R_S\left|\frac{d I(R_S)}{dR_S}\right|}{1+I(R_S)} \right) = T R_S \frac{1}{1+I(R_S)}\, ,
\end{equation}
where the last equality is obtained using Eq.~(\ref{trsgen}).
 Eq.~(\ref{trs}) then  implies that $\frac{d\ln R_S}{d\ln_M}=1$.

Hence we only need to prove that the product $M R_S$ is maximal for Einstein theory. This follows immediately from Eq.(\ref{mrs}): since for a consistent theory $I(R_S)$ is positive, the maximal value for $M$ is reached when $I(R_S)=0$, namely in Einstein's theory.

We have thus shown that the length scale  $l_{SCBH}$ is the shortest scale below which semiclassical BHs do not exist also in an arbitrary consistent extension of Einstein's gravity theory. Combining this with the argument leading to (\ref{UVbound}) completes the proof of our claim.

\section{Examples}

In this section we present two examples of generalized theories of gravity. The first is Einstein gravity in a compactified higher-dimensional spacetime and the second is weakly coupled string theory. In the two examples  the microscopic description of the theories as well as the effective description are both known so our assumptions and results can be  explicitly checked and verified.

\subsection{Einstein Gravity in higher dimensions}
\label{extraD}

Let us consider a $(4+n)$-dimensional Einstein gravity theory, with the $n$ extra dimensions compactified  on a torus  of radius  $R$. We shall first consider the case when there are no branes
that could violate the translation invariance in extra coordinates.

The  microscopic theory is characterized by its $(4+n)$-dimensional Planck length $l_{4+n}$ and we assume that it has $N_{4+n}$ species. The  four-dimensional (4D) theory has its standard four-dimensional Planck length $l_P$ and contains  $N_4$ species whose relation to $N_{4+n}$ we will determine shortly. For distances $r>R$ the theory behaves as a 4D Einstein gravity with small corrections (that we shall ignore). For $r<R$ the theory is essentially a $(4+n)$-dimensional theory, which, from a 4D point of view, deviates substantially from 4D Einstein gravity.
In  this higher-dimensional model   the analogue of (\ref{defl*}) reads:
\begin{equation}
\label{uvhd}
l_{SCBH}=l_{4+n} (N_{4+n})^{1/(n+2)}.
\end{equation}
This scale $l_{SCBH}$ is also compatible with our definition of the cutoff $l_{UV}$, since gravitational interaction among elementary particles remains weak for all scales  $l\ge l_{SCBH}$ as we have shown in Sect.~(\ref{EGefflambda}). We can therefore take any $l_{UV} \ge l_{SCBH} $, but, for the sake of definiteness, we shall simply assume that $l_{UV} = l_{SCBH} $.

As observed  in \cite{Dvali},  this setup provides an explicit example for verifying bound (\ref{bound}).
Indeed,  geometrically, there is a well-known relation,
\begin{equation}
M_P^2 = M_{4+n}^2 ( M_{4+n}R)^{n} \, ,
\label{scales}
\end{equation}
where $M_{4+n}=1/l_{4+n}$ is the $4+n$ dimensional Planck mass.
From the 4D point of view, the factor $(M_{4+n}R)^{n}$ is the number of Kaluza-Klein modes per each higher-dimensional species
\begin{equation}
\label{nhd}
\left(\frac{R}{l_{4+n}}\right)^{n} = \frac{N_4}{N_{4+n}}.
\end{equation}
The effective gravitational coupling in $4+n$ dimensions $\lambda_{G,n}$ is given by
\begin{equation}
\label{lambdaHD}
\lambda_{G,n}(l) = N_{4+n} \left(\frac{l_{4+n}}{l}\right)^{n+2}
\end{equation}
which reaches unity at the scale $l_{UV}$ defined in Eq.~(\ref{uvhd}). Obviously, below this scale $\lambda_{G,n}(l)<1$, in particular $ \lambda_{G,n}(R)< \lambda_{G}(R)$.
\begin{equation}
\label{lambdaHDR}
\lambda_{G,n}(R) =N_4 \frac{l_P^2}{R^2} \left(\frac{l_{4+n}}{R}\right)^{n}=\lambda_G(R) \left(\frac{l_{4+n}}{R}\right)^{n}
\end{equation}

Let us turn now to the check of our assumptions and results about the properties of BHs in this example.
In the microscopic theory the BH metric is:

$
ds^2=-\left(1-\left(R_S/r\right)^{n+1}\right)dt^2+\frac{1}{1-\left(R_S/r\right)^{n+1}} dr^2 + r^2 d\Omega_{2+n}^2.
$

From this metric we see that $T R_S= (n+1)/4\pi$. Ignoring as usual numerical factors this relation satisfies Eq.~(\ref{trs}).
Concerning the dependence of the BH mass $M$ on the Schwarzschild radius $R_S$ let us first discuss the region $R_S>R$. We may calculate the (ADM) mass of the BH  in the microscopic theory as
$
M=M_{4+n}^2 R_S (M_{4+n} R)^n.
$
From the 4D point of view, using Eq.~(\ref{scales}), we find
$
M=M_{P}^2 R_S
$
i.e. exactly the standard result for a 4D Einstein theory.
On the other hand, repeating the same procedure  for  $R_S<R$ we recover the well-known result \cite{myersperry}
$R_S=\frac{1}{\sqrt{\pi}}\frac{1}{M_{4+n}}\left(\frac{M}{M_{4+n}}\right)^{1/(n+1)} \left(\frac{8\Gamma((n+3)/2)}{n+2}\right)^{1/(n+1)}$, which, dropping numerical factors, turns into $M=M_{4+n}^2 R_S (M_{4+n} R_S)^n$. From the 4D point of view the resulting $M(R_S)$ is
\begin{equation}
\label{mrsnone}
M=M_P^2 R_S \left(\frac{R_S}{ R}\right)^n.
\end{equation}
Here we can see explicitly that for the generalized theory the mass of a BH of radius $R_S$ is smaller than the corresponding one in Einstein's  theory
$M<M_E=M_{P}^2 R_S$. We can also verify that $d\ln M /d\ln R_S = n+1$ which, using the expression for the temperature, becomes $d\ln M /d\ln R_S =4\pi T R_S$.

A non-trivial subtlety appears for compactification on manifolds that are not translation invariant in the compact  dimensions \cite{oriol}, e.g. when space includes branes with localized species. In such a case,  one seems to find  a contradiction with the assumption~(\ref{BB}) about the universal thermal evaporation of semiclassical BHs. The BHs that are {\it not}  pierced by a given brane cannot evaporate into the species localized on that brane,  due to locality in the compact dimensions.  This would naively suggest that there can be semiclassical BHs that do not evaporate  democratically, in sharp  contradiction with our assumptions. The resolution of this apparent conflict can be found by noticing that such BHs are unavoidably time-dependent at the classical level. As shown in \cite{oriol}, BHs that evaporate non-democratically cannot be classically static and evolve in time until the partial evaporation rates into all the species equalize.  This ``democratization'' process restores consistency with our assumption (\ref{BB}).

\subsection{Weakly coupled string theory}

In string theory, the UV scale is the string length  $l_{UV}=l_s$.  In weakly coupled string theory the well known relation between the string length and the Planck length $l_P=l_s g_s$ is expressed in terms of the string coupling constant $g_s$. So in this case $l_{SCBH}=l_P \sqrt{N}=l_s g_s\sqrt{N}$, where $N$ is the number of massless string excitations. Inequality~(\ref{UVbound}) implies then that
\begin{equation}
\label{gsn}
g_s^2 N<1.
\end{equation}
The effective gravitational coupling $\lambda_G$ is given by
\begin{equation}
\label{lgstring}
\lambda_G = N l_P^2/l_s^2= Ng_s^2
\end{equation}
and the bound~(\ref{bound}) also implies the inequality~(\ref{gsn}). Such inequality defines what we should really call ``weakly-coupled" string theory. Let us remark that for very small string coupling  the mass of a string whose size is $l_s$ is given by $M_c = M_s g_s^{-2} = M_P g_s^{-1}$. Such a BH lies on the so-called ``correspondence" curve \cite{corr}  between fundamental strings and BH and indeed its entropy $S = S_c = g_s^{-2}$ can be computed either by the string  or by the BH-entropy formula.
Here we wish to emphasize that weakly coupled string theory is an example of a theory that contains semiclassical BH with sizes all the way down to the cutoff scale $l_s$ but where, at the same time, BHs remain as classical as one wishes for all length scales since $M R_S =  g_s^{-2} > N$. Then, supposedly, they stop existing as BHs and turn into ordinary weakly-coupled strings .

Supposedly in string theory there are no BHs smaller than $l_s$, since at that point BHs turn into ordinary, non collapsed, ``large"  objects \cite{Bowick}.
More exotic possibilities can be considered, however, where BHs smaller than $l_s$ and with temperature higher than the Hagedorn temperature $M_s$ might exist.
We have checked that, at least for BHs of temperature smaller than $T_*\equiv \sqrt{M_PM_s}$, our bound still makes sense.


\section{Some physical consequences of the bound}

We shall now discuss some physical consequences of the bounds derived in the previous section and briefly mention previous discussions about the relevance of our bound.
Several applications of the bound (\ref{bound})  (e.g.,  for hierarchy problem, cosmology  and physics of micro black holes) where already discussed in \cite{Dvali,cesar}  (and references therein), and will not be repeated here.

\begin{itemize}

\item{{\bf Triviality of quantum gravity}}

By this we mean, in the standard sense of the word triviality (as in  $\lambda \phi^4$ quantum field theory), that $G_N\to 0$ in the infinite energy cutoff limit, or equivalently in the $l_{UV}\to 0$ limit.  This result follows immediately from (\ref{bound}): $\lambda_G(l_{UV})= N(l_{UV}) G_N/l_{UV}^2 <1$ implies that
\begin{equation}
\label{GNbound}
G_N<   \frac{l_{UV}^2}{N(l_{UV})}.
\end{equation}
Consequently, any attempt to renormalize a consistent extension of Einstein's Gravity with a finite fixed number of light stable species (including, for instance, ${\cal N}=8$ supergravity) will fail because the removal of the cutoff necessarily will make gravity trivial in the infrared.

\item{{\bf The Sakharov induced gravity limit for a finite UV cutoff}}

In Sakharov's induced-gravity limit the tree-level value of Newton's constant is taken to infinity, thus removing the Einstein-Hilbert term from the tree-level action. A concrete example is string theory in the infinite-string-coupling limit where also the tree-level kinetic term of the gauge fields vanishes. Inequality~(\ref{GNbound}),
which applies to the physical renormalized coupling, implies that, in this limit, the {\it renormalized} Newton constant will remain finite and bounded.

\item{{\bf String theory}}

We have already discussed how weakly coupled string theory satisfies our bound. What about moderately or strongly coupled string theory?
Such a situation is defined by having a positive (and possibly large) VEV for the dilaton $\phi$, since $g_s \sim e^{\phi}$.
One possibility is that the bound never gets saturated for all scales larger than $l_{s}$. For example,  in some string-theory backgrounds it is known that the infinite string coupling limit of the theory corresponds to the zero-coupling limit of another string theory.

An appealing  alternative is that the bound gets saturated either at some finite value $g_*$ of $g_s$ or as $g_s \rightarrow \infty$:
\begin{equation}
\label{GNsat}
\lambda_G(g_*) = \frac{NG_N(g_*)}{l_s^2} \rightarrow  1.
\end{equation}

If $N$ is in the hundreds or thousands (as in the case of large unified gauge groups),  Eq. (\ref{GNsat}) could provide an interesting value for the ratio $\l_s/l_P$ by making  $M_s$ approach the GUT scale  of $\sim 10^{16} {\rm GeV}$. This can be contrasted with the perturbative situation in which:
\begin{equation}
\label{GNpert}
\frac{G_N}{l_s^2} \sim \alpha_{GUT}\, ,
\end{equation}
giving, typically, $M_s \sim 10^{17} {\rm GeV}$. It is clear that (\ref{GNsat}) agrees with (\ref{GNpert}) if $ \alpha_{GUT}\sim 1/N$.
However, the 't Hooft coupling of the unified gauge theory is given by $\lambda_{GUT} = \alpha_{GUT} \widetilde{N}$ where $\widetilde{N}$ is  of order of the rank (or the quadratic Casimir) of the gauge group.  In general we expect $N \sim \widetilde{N}^2 \gg \widetilde{N}$ since $N$ is the total number of light species which is roughly the number of gauge bosons. Assuming that  also $\lambda_{GUT}$ saturates in the strong coupling limit, our bound would allow to lower the ratio $M_s/M_P$ relative to its perturbative value \cite{GV01}.

Also for the strong-coupling limit more exotic possibilities exist.
Of course,  the bound (\ref{bound}) is based on the existence of degrees of freedom  that fall within our definition of species,  implying that they must be weakly-coupled at least
within some finite energy interval.  If there exists a sensible limit of a strongly coupled string theory allowing for such an interval,  then the bound applies. This could be the case if, for example,
the strongly  coupled string theory allows a mass gap with the lowest lying string zero modes
being weakly coupled within that gap.   The bound (\ref{bound}) then would relate the width of that
gap to the number of zero modes.

\item{\bf Entropy bounds}

In Einstein gravity with a fixed number of species the bound (\ref{Ebound}) implies a bound on the Bekenstein-Hawking entropy of BHs $S_{BH}(R_S)=M_P^2 R_S^2$,
\begin{equation}
\label{Sbound}
S_{BH}>N.
\end{equation}
It is quite likely that one can argue directly in favor of (\ref{Sbound}) by using arguments that rely on the generalized second law of thermodynamics \cite{gsl} and hence that the validity of (\ref{Sbound}) is more general. A saturation of the bound at some finite scale $S_{BH}(R_*)=N$ is quite interesting since it may imply that the origin of BH entropy is entirely from the matter sector.

\end{itemize}

\section*{Acknowledgements}
RB thanks the theory division, CERN for hospitality. The research of RB is supported by ISF grant 470/06 and by PICS grant.  The research of GD  is supported in part  by European Commission  under
the ERC advanced grant 226371,  by  David and Lucile  Packard Foundation Fellowship for  Science and Engineering and  by the NSF grant PHY-0758032.

\end{document}